\documentstyle[aps,prd,tighten,epsf]{revtex}
\begin{document}
\draft
\newcommand{\be}{\begin{equation}}
\newcommand{\ee}{\end{equation}}
\newcommand{\ben}{\begin{eqnarray}}
\newcommand{\een}{\end{eqnarray}}

\newcommand{\la}{{\lambda}}
\newcommand{\La}{{\Lambda}}
\newcommand{\Om}{{\Omega}}
\newcommand{\ta}{{\tilde a}}
\newcommand{\bg}{{\bar g}}
\newcommand{\bh}{{\bar h}}
\newcommand{\si}{{\sigma}}
\newcommand{\th}{{\theta}}
\newcommand{\C}{{\cal C}}
\newcommand{\D}{{\cal D}}
\newcommand{\cA}{{\cal A}}
\newcommand{\cT}{{\cal T}}
\newcommand{\cO}{{\cal O}}
\newcommand{\eeo}{\cO ({1 \over E})}
\newcommand{\G}{{\cal G}}
\newcommand{\cL}{{\cal L}}
\newcommand{\T}{{\cal T}}
\newcommand{\M}{{\cal M}}

\newcommand{\p}{\partial}
\newcommand{\na}{\nabla}
\newcommand{\ssum}{\sum\limits_{i = 1}^3}
\newcommand{\dssum}{\sum\limits_{i = 1}^2}
\newcommand{\tal}{{\tilde \alpha}}

\newcommand{\tp}{{\tilde \phi}}
\newcommand{\tPhi}{\tilde \Phi}
\newcommand{\tpsi}{\tilde \psi}
\newcommand{\tim}{{\tilde \mu}}
\newcommand{\tr}{{\tilde \rho}}
\newcommand{\tir}{{\tilde r}}
\newcommand{\rp}{r_{+}}
\newcommand{\hr}{{\hat r}}
\newcommand{\rv}{{r_{v}}}
\newcommand{\dr}{{d \over d \hr}}
\newcommand{\hhf}{{\hat \phi}}
\newcommand{\hhM}{{\hat M}}
\newcommand{\hhQ}{{\hat Q}}
\newcommand{\hht}{{\hat t}}
\newcommand{\hhr}{{\hat r}}
\newcommand{\hhS}{{\hat \Sigma}}
\newcommand{\hhD}{{\hat \Delta}}
\newcommand{\hhm}{{\hat \mu}}
\newcommand{\hro}{{\hat \rho}}
\newcommand{\hhz}{{\hat z}}

\newcommand{\tD}{{\tilde D}}
\newcommand{\tB}{{\tilde B}}
\newcommand{\tV}{{\tilde V}}
\newcommand{\tsig}{{\tilde \sigma}}
\newcommand{\hT}{\hat T}
\newcommand{\tF}{\tilde F}
\newcommand{\tT}{\tilde T}
\newcommand{\hC}{\hat C}
\newcommand{\ep}{\epsilon}
\newcommand{\bep}{\bar \epsilon}
\newcommand{\ppp}{\varphi}
\newcommand{\Ga}{\Gamma}
\newcommand{\ga}{\gamma}
\newcommand{\hth}{\hat \theta}
\title{Cosmological Black Holes on Branes}

\author{Marek Rogatko}

\address{Institute of Physics \protect \\
Maria Curie-Sklodowska University \protect \\
20-031 Lublin, pl.Marii Curie-Sklodowskiej 1, Poland \protect \\
rogat@tytan.umcs.lublin.pl \protect \\
rogat@kft.umcs.lublin.pl}

\date{\today}
\maketitle
\smallskip
\bigskip
\begin{abstract}
We examined analytically a cosmological black hole domain wall system.
Using the C-metric construction we derive the metric for the spacetime describing an
infinitely thin wall intersecting a cosmological black hole. We studied the 
behaviour of the scalar field describing a self-interacting cosmological
domain wall and find the approximated solution valid for large distances.
The {\it thin wall} approximation and the
back reaction problem were elaborated finding that the topological kink
solution smoothed out singular behaviour of the zero thickness wall
using a core topological and hence thick domain wall. We also analyze the
nucleation of cosmological black holes on and in the presence of a domain walls
and conclude that the domain wall will nucleate small black holes on it
rather than large ones inside.
\end{abstract}
\baselineskip=18pt
\par
\section{Introduction}
There has been recently a considerable interest in topological defects arising during phase
transitions in the early universe which can play an important role in our
understanding of cosmological evolution of its own right \cite{vil94}.
\par
Domain walls are subject of the intense studies because of the concept
that our universe might be a brane or a defect emerged in some higher dimensional
geometry. This idea appears in the recent unifications attempts such
as superstring theories or M-theory \cite{hor96,luc99}. 
As was shown in \cite{hor96}
the $E_{8} \times E_{8}$ heterotic string theory at strong
coupling may be described in terms of M-theory acting in eleven-dimensional
spacetime with boundaries where ten-dimensional supersymmetric
Yang-Mills gauge degrees of freedom reside on two boundaries, {\it
branes at the end of the world}. In contrast to the gauge fields
acting only in a lower-dimensional submanifold gravitation can penetrate
the bulk.
One also hopes that this idea
will be helpful to investigate the very intriguing problem of hierarchy,
i.e., the huge gap between the electroweak scale and the Planck scale.
Randal and Sundrum \cite{ran99} proved that for nonfactorizable geometries in
five-dimensional spacetime the zero modes of the Kaluza-Klein dimensional
reduction can be localized in four-dimensional submanifold. 
They also managed to reproduce the Newtonian gravity at large distances.
\par
Interests of cosmic string black hole systems were motivated, among
other things, by the possible enlargement of the {\it no-hair} \cite{bl}
conjecture to the problem of the nontrivial topology of field
configurations surrounding black holes. In Ref.\cite{dow92} the numerical
and analytical evidence for Abelian Higgs vortex acting as a long
hair for Schwarzschild black hole or Reissner-Nordstr\"om (RN) solution were
provided. Mann and collaborators investigated the problem of vortex in the spacetime of
cosmological black holes \cite{man02} while
the problem of the vortex solutions in the background of an electrically charged black string
was examined in \cite{deg02}, where it was shown that black string could also support
the Abelian Higgs field as hair.\\
On the other hand, in dilaton gravity 
being the low-energy limit of the superstring theories,
the problem of Abelian Higgs vortex 
acting as hair on the dilaton black holes
was studied in Refs.\cite{string,san00}. When the dilaton black hole 
tends to extremality one has always expulsion
of the flux from the black hole (the so-called {\it Meissner effect}).
This phenomenon was justified both by numerical and analytical investigations.
For the vortex which is thick enough the black hole will expel its flux, causing
the field to remain in its symmetric state on the event horizon. It will be not amiss
to expect the similar phenomenon to occur in the case of the domain wall black hole system.
\par
The problem of domain wall and black hole in general relativity was
studied in \cite{emp01} and in dilaton gravity in Ref.\cite{rog01}. 
Among all, it was shown how 
to smooth this singular solution with a thick domain wall. The process of
nucleation of black holes on domain walls was also tackled. In the case
of dilaton gravity it was revealed analytically that the extreme dilaton black hole
always expelled domain wall (we have to do with the so-called {\it Meissner effect}).
The numerical studies of a topological domain wall and Schwarzschild black hole
were conducted in \cite{mor00,mor03}. In dilaton gravity all the analytical outcomes were
confirmed numerically in \cite{mod03}.
\par
Our paper is essentially concerned with the problem of a domain wall and a
cosmological black hole system, providing some continuity with the previous works
\cite{string,rog01,mod03}.
The paper is organized in the following way. In Sec.II
we derive the infinitisemal domain wall cosmological black hole metric.
Sec.III is devoted to the problem of self-gravitating cosmological domain wall.
We find the approximated solution for scalar field equations describing the domain wall
valid for large distances. In Sec.IV we examined the scalar field equations in the
background of RN de Sitter (RN-dS) and RN-AdS spacetimes
as well as in the cosmological C-metric background. The {\it thin wall}
approximation was also described. We also tackle 
the problem of gravitational
back-reaction. We justify that as in general relativity and dilaton gravity
also in the cosmological background a topological kink smoothes out a 
shell-like singularity of a domain wall. 
In Sec.V we tackle the problem of the expulsion of domain walls from
the black holes and find the approximate form for the scalar field to
expulsion place place. 
In Sec.VI we study the production of cosmological black holes on the domain walls
and nucleation of pairs of static cosmological black holes in the presence of
a domain wall.
We summarize and conclude our
investigations in Sec.VII.

\section{Black holes with cosmological constant}
In this section we derive solution describing an infinitisemally
thin domain wall with a cosmological black hole as the static spherically
symmetric solution of Einstein equations.
We begin with
the action of the considered theory in the form as follows:
\be
S = \int d^4 x \sqrt{- g} \bigg[ R - 2 \La - F_{\mu \nu}F^{\mu \nu} \bigg],
\ee
while equations of motion derived from the variational principle imply
\ben
\na_{\mu} F^{\mu \nu} = 0, \\ \label{q}  
G_{\mu \nu} + \La g_{\mu \nu} = T_{\mu \nu}(F),
\een
where the energy-momentum tensor yields
\be
 T_{\mu \nu}( F) =  4 F_{\mu \rho}
F_{\nu}{}{}^{\rho} - g_{\mu \nu}F^2.
\ee
In general
C-metric solution has interpretation as a two oppositely charged
black holes undergoing acceleration.
The various aspects of the C-metric were vastly studied in literature
\cite{kin70}-\cite{dia03b}.
\par
In our studies we shall use the
C-metric for cosmological black hole is as follows:
\be
ds^2 = {1 \over A^2 \big( x - y \big)^2}
\bigg[ H(y) dt^2 - {dy^2 \over H(y)} + {dx^2 \over G(x)} + G(x) d\phi^2 \bigg],
\label{coc}
\ee
where 
\ben
G(x) &=& a - {\La \over 3 A^2} - b x^2 - 2 m A x^3 - q^2 A^2 x^4, \\ 
H(y) &=& a - b y^2 - 2 m A y^3 - q^2 A^2 y^4,
\een
$\La = {3k \over l^2}$, $k= \pm 1$ is the cosmological constant and $A$ is the acceleration
parameter. The $U(1)$ gauge field in the magnetic and electric case are respectively
$A_{\phi} = - qx$ and $A_{t} = qy$.\\
The freedom of a redefinition of the parameters appearing in C-metric
enables one to map C-metric onto another C-metric
up to a conformal transformation \cite{kin70}. It can be typically eliminated
by setting $a = 1$ and $b = 1$.\\
The roots of $G(x)$ are denoted by $x_{1}, x_{2}, x_{3}, x_{4}$ in ascending order, while the 
roots of $H(y)$ are written as $y_{1}, y_{2}, y_{3}, y_{4}$. The surface $y = y_{1}$ is the inner
black hole horizon, $y = y_{2}$ is the outer black hole horizon while $y = y_{3}$ is
interpreted as the accelerated or cosmological horizon. The largest root has no physical
meaning.
\par
In order to study the cosmological constant domain wall black hole we shall 
use the method described in \cite{emp00a,emp00b}.
According to
the Israel
procedure \cite{isr66}
the discontinuity of the extrinsic curvature is provided by the
tension $\tsig$ of a domain wall in the cosmological background.
Then, it implies the following:
\be
\big[ K_{ij} \big] = 4 \pi G \tsig h_{ij},
\ee
where $h_{ij}$ is the metric induced on the wall. 
Having in mind \cite{emp01}, an
appropriate umbilic surface can be found at $x = 0$. The surface has
normal $n = {1 \over Ay} dx$, and the induced metric may be written as
\be
ds^2 = {1 \over A^2 y^2} \bigg[ H(y) dt^2 - {dy^2 \over H(y)} + 
\bigg( 1 - {\La \over 3 A^2} \bigg) d\phi^2 \bigg].
\ee
A short calculations reveal that 
the extrinsic curvature in this case is equal to $K_{ij} = A h_{ij}$ where the
Israel condition implies that the domain wall tension in the cosmological 
background is equal to $
\tsig = \sigma + \La = 
A/ 2 \pi G$.
After changing the variables $r = - 1/Ay$ and $T = t/A$, the metric
reduces to the form as follows:
\be
ds^2 = - \bigg( 1 - {2 m \over r} + {q^2 \over r^2} - A^2 r^2 \bigg) dT^2 +
{dr^2 \over \bigg( 1 - {2 m \over r} + {q^2 \over r^2} - A^2 r^2 \bigg)}
+ r^2 \bigg( 1 - {\La \over 3 A^2} \bigg) d\phi^2.
\ee
As in Ref.\cite{emp01} we decide the conical singularity to lie at $x = y_{3}$
on the side where $x<0$. Consequently, under this assumption if we build the domain
wall black hole system by gluing two copies of each side $x>0$, we shall get rid of
the string from the spacetime.
\par 
The charge of the considered black hole can be measured by means of the integration
of the flux conducted through the sphere surrounding it. It results in the form
\be
Q_{BH-dw} = 2 {\Delta \phi \over 4 \pi} 
\bigg[ A_{\phi}(x = y_{4}) - A_{\phi}(x = 0) \bigg] =
{ 2 y_{4} \over q A^2 (y_{4} - y_{1})(y_{4} - y_{2})(y_{4} - y_{3})}.
\label{lad}
\ee
The factor two in Eq.(\ref{lad}) appears because we should take into account both
sides of the domain wall. The black hole built on the domain wall will neither
swallow up the brane nor slide it off \cite{emp00} due to the fact of existing an
elastic force action exerted by the brane on the cosmological black hole.
\par
Thus, using the C-metric construction \cite{emp00a,emp00b} we have derived
an equivalent of the thin cosmic string black hole solution \cite{afv}, namely
the metric of infinitisemally thin domain wall bisecting a cosmological
black hole.

\section{Self-gravitating cosmological domain wall}
In this section our considerations will be devoted to 
a general matter Lagrangian with real Higgs field and the symmetry
breaking potential.
The Lagrangian density for the scalar field can be written as 
\be
{\cal L}_{dw} = - {1 \over 2} \na_{\mu} \varphi \na^{\mu} \varphi - U(\varphi).
\ee
We assume that 
the symmetry breaking potential $U(\varphi)$ has a discrete set of degenerate minima.
The energy-momentum tensor for the scalar fields responsible for the domain wall formation 
implies the following:
\be
T_{ij}(\varphi) = - {1 \over 2} g_{ij} \na_{m} \varphi \na^{m} \varphi
- U(\varphi) g_{ij} + \na_{i} \varphi \na_{j} \varphi.
\ee
As usual, 
for the convenience, we scale out parameters via transformation $ X = {\varphi / \eta}$
and $\ep = 8 \pi G \eta^2$. 
It happened that 
the parameter $\ep$ represents the gravitational 
strength and is connected with the gravitational interaction of the Higgs field.
Let us define
$V(X) = {U(\varphi) \over V_{F}}$, where $V_{F} = \lambda \eta^4$. Then, we arrive
at the relation:
\be
8 \pi G {\cal L}_{dw} = - {\ep \over w^2} \bigg[
w^2 {\na_{\mu} X \na^{\mu} X \over 2} + V(X) \bigg],
\label{dww}
\ee
where
 $w = \sqrt{{\ep \over 8 \pi G V_{F}}}$
is
the inverse mass of the scalar  after symmetry breaking, 
which also characterize the width of the domain wall defect.
Using Eq.(\ref{dww}) we can derive the relation for $X$ field. Namely, it may be written as
\be
\na_{\mu} \na^{\mu} X - {1 \over w^2}{ \p V \over \p X} = 0,
\label{xx}
\ee
where $V(X) = {1 \over 4}(X^2 - 1)^2$.
In our next step we try to determine the asymptotic behaviour of $X$
in the spacetime with cosmological constant. We shall not specify the exact sign of
cosmological constant. Therefore our solution will depend on a constant $k = \pm 1$.
The metric under consideration has the form:
\be
ds^2 = - \bigg( 1 - {\La \over 3}r^2 \bigg) dt^2 +
{dr^2 \over \bigg( 1 - {\La \over 3}r^2 \bigg)} + 
 r^2 (d \theta^2
+ \sin^2 \theta  d \phi^2),
\ee
where $\La = 3 k/l^2$.
One can verifies that the Eq. of motion for the scalar field (\ref{xx}) may be written as
\be
\bigg( 1 - {k r^2 \over l^2} \bigg) X_{,rr} +
{2 \over r} \bigg( 1 - {2 k r^2 \over l^2} \bigg) X_{,r} +
{1 \over r^2} X_{,\theta \theta} + {1 \over r^2} \cot \theta X_{,\theta} -
{1 \over w^2}{\p V \over \p X} = 0.
\ee
Next, after changing the variables $z = r \cos \theta$ we get
\be
\bigg( 1 - {k z^2 \over l^2} \bigg) X_{,zz}
 - {4 k z \over l^2} X_{,z}
 - {1 \over w^2} X (X^2 - 1) = 0.
\label{xxz}
\ee 
The exact analytic solution of Eq.(\ref{xxz}) is not known.
However if one assumes that $X( z \rightarrow \infty ) \rightarrow 1$,
i.e., its value tends to the value of the scalar field at large distances,
we are able to examine the behaviour of its magnitude.
The inspection of the equation of motion reveals that this relation
is approximately satisfied by $X = X_{0} \simeq 1$
for large $z$, where
$X_{0}$ is the the minimum of the considered potential $V(X)$. This minimum 
is the vacuum value of the scalar field configuration. Let us assume further, that the
fluctuations about the considered vacuum value is $\xi(z)$,
namely
\be
X \simeq X_{0} + \xi(z).
\ee
Then it implies the following:
\be
{k z^2 \over l^2} {d^2 \xi \over d z^2} + {4 k z \over l^2} 
{d \xi \over dz}
+ {1 \over w^2} \xi (z) \simeq 0.
\label{app1}
\ee
In the derivation of Eq.(\ref{app1}) we have neglected terms of order of unity
in the coefficients of the first and second terms. The approximated
solution for large $z$ has the following form:
\be
X_{(s-gr)} = X_{0} \bigg[ 1 - \bigg( {z \over z_{0}} 
\bigg)^{- 3 + \sqrt{9 - 4 \beta^2 }\over 2} \bigg],
\ee
where we have denoted by $\beta^{2} = l^2/ w^2 k$ and $z_{0}$ is a constant of
integration.

\section{Domain walls in the cosmological context}
In this section we first demonstrate how the distributional domain wall with a cosmological
black hole on it can be consistently find as a limit of physical,
field theoretical topological defects. 
Let us consider the line element of RN-dS and RN-AdS metric
which can be expressed as follows:
\be
ds^2 = - \left ( 1 - {2 M \over r} + {Q^2 \over r^2} -{ \Lambda \over 3} r^2  
\right ) dt^2 +
{d r^2 \over {\left ( 1 - {2 M \over r} + {Q^2 \over r^2} -{ \Lambda \over 3} r^2
\right )}} + r^2 (d \theta^2
+ \sin^2 \theta  d \phi^2),
\ee
For the black hole vortex system the scalar fields were approximated as
a functions of $\sqrt{g_{\phi \phi}}$ \cite{dow92}.
Guessing the ansatz for the scalar field in the form $X(z) = X(r \cos \theta)$ we get
the expression:
\ben
\na_{\mu} \na^{\mu} X &=&
X_{,zz} \bigg[
1 - {2 M z^2 \over r^3} + {Q^2 z^2 \over r^4} - {\La \over 3} z^2 \bigg]
+ X_{,z} \bigg[
{Q^2 z \over r^4} - {4 M z \over r^3} - {5 \over 3}\La z + {1 \over z}
\bigg] \\ \nonumber
&=&
X_{,zz} - {\La \over 3} z^2 X_{,zz} - {5 \over 3} \La z X_{,z} + {\cal O }({1 \over M}).
\een
We remark that $r$ is far more greater than $M$ outside the horizon of 
the black hole. Then, if the thickness of the domain wall is much less
smaller than the black hole horizon, i.e., $M \gg 1$ one can draw a conclusion 
that the scalar field $X$ tends to its self-gravitating solution
in AdS or dS spacetime depending on the sign of the cosmological constant.
It can be observed that a thin wall can be painted on the cosmological black hole.
This fact was also confirmed for RN and dilaton black hole solution in Refs.
\cite{emp01,rog01,mod03}. Our preliminary numerical
studies \cite{mod04} reveals that this situation also takes place for the 
cosmological black holes.
\par
Because we expect that the gravitating system under consideration will have
C-metric form, one can 
see that for the variable $x$ is responsible $\cos \theta$ and this fact enables us
to guess that $z = x/Ay$. 
Consequently under this assumption,
one can write
\ben
\na_{\mu} \na^{\mu} X &=& X_{,zz}
\bigg( z A - 1 \bigg)^2 \bigg[ G(x) - A^2 z^2 H(y) \bigg] +
X_{,z} (z A - 1)^2 \bigg[
G_{,z}(x) - z^2 A^2 H(y)_{,z} \bigg] \\ \nonumber
&+&
X_{,z} G(x) (z A - 1) \bigg[ - 2 A - 2 {1 \over z } + 
\bigg( {z A - 1 \over z^2} \bigg) \bigg] \\ \nonumber
&+& X_{,z} H(y) z A (z A - 1) \bigg[
- A^2 z + 5 A \bigg].
\een
Using the {\it thin wall} approximation \cite{emp01} i.e., 
we expect the black hole horizon is to be large ($A \mid y_{2} \mid \ll 1$) and
considering the regime of large acceleration radius
($A \mid y_{3} \mid \ll 1$), one concluded that 
\be
\na_{\mu} \na^{\mu} X =
X_{,zz} - {\La \over 3} z^2 X_{,zz} - {5 \over 3} \La z X_{,z} + {\cal O }(A^2).
\ee
Thus, we have arrived at the conclusion that $X = X_{(s-gr)}$ is a good approximation solution
of the field equations in the cosmological $C$-metric spacetime.
\par
The other interesting problem we shall
take up a problem of the gravitational back reaction
for the thick domain wall and cosmological black holes described by the spacetime metric
(\ref{coc}). As in Ref.\cite{emp01,rog01} we 
denote for simplicity $\Omega = A(x -y)$ and we
use linearized calculations in
$\ep = 3/2A$, writing $\Omega = \Omega_{0} + A \Omega_{1}$ and so on.
In the vicinity of the core of the domain wall we have $\Omega_{1}/ \Omega_{0}
= {\cal O}(1)$, while in the region far away from the core it tends to zero.
\par
From Eqs. of motion for gauge field, namely $\p_{x} [ \p_{x} A_{\phi} ] = 0$
and the form of the first order perturbed solution determined by
$A_{\mu}^{(1)} = f(z) A_{\mu}^{(0)}$ one has that $f(z)$ is constant.
Thus, the gauge field potential is unaltered by the presence of the domain wall.
\par
As was remarked in \cite{emp01,rog01} since the fact that variation
of the extrinsic curvature due to the domain wall is carried by $\Omega$,
we guess that $G(x)$ and $H(y)$ will take effectively their background
values. In this case, 
after some lengthy calculations one reaches to the following equations of motion:
\ben
R_{\phi}{}{}^{\phi} - R_{x}{}{}^{x} &=&
- 2 G(x) A^2 y \bigg( A z - 1 \bigg) f_{,xx} = - \ep \bigg(
X_{(s-gr),z} \bigg)^2, \\ 
R_{t}{}{}^{t} - R_{y}{}{}^{y} &=&
2 H(y) A^2 y \bigg( A z - 1 \bigg) f_{,yy} = {\cal O}(A^2), \\
R_{t}{}{}^{t} - R_{\phi}{}{}^{\phi} &=&
{\Omega \over 2}
\bigg[ \Omega G(x)'' + \Omega H(y)'' - 2 \bigg(
H(y)' \Omega \Omega_{,y} + G(x)' \Omega \Omega_{,y} \bigg)
\bigg] \\ \nonumber
&=& {Ay \over 2} \bigg( A z - 1 \bigg)^2 
\bigg[ G(x)'' + H(y)'' \bigg] - A^3 y^2 \bigg( A z - 1 \bigg)^2
\bigg[ H(y)' (f_{,y} - 1) + G(x)' f_{,x} \bigg]
= - 4 q^2 \Omega^4.
\een
As in Ref.\cite{emp01} the relations for Ricci tensors suggest that $\Omega$ may be presented
as follows:
\be
\Omega = A (f - y),
\ee
where $f_{0} = \mid x \mid$.\\
Because of the fact that $f_{,x} = {\cal O}(A)$ and 
$f_{,y} = {\cal O}(A^2)$ we draw a conclusion
that the Einstein's equations in the cosmological context will be 
satisfied to the leading order
in $A$. Substituting the above ansatz for $\Omega$ 
and the ansatz for the scalar field as a solution
for self-gravitating scalar field in the spacetime with non-vanishing cosmological constant,
we arrive at the following expression:
\be
f(x) = - \ep A^2 y^2 X_{0}^2 {(-5 + \sqrt{9 - 4 \beta^2})^2 \over 
4 (-4 + \sqrt{9 - 4 \beta^2}) (-3 + \sqrt{9 - 4 \beta^2})}
\bigg({ x \over A y z_{0}} \bigg)^{-3 + \sqrt{9 - 4 \beta^2}}.
\ee
From the above back reaction considerations
one can draw a conclusion
that as in the domain wall black hole system in general relativity \cite{emp01},
and domain wall dilaton black hole system in the low energy string theory
\cite{rog01} in the case of cosmological black hole domain wall system a
kink solution also smoothes out the shell-like singularity of the infinitisemal
domain wall.

\section{Expulsion of the domain wall by thick brane}
\subsection{General considerations}
In this section we shall confine our attention to the problem of expelling thick domain 
walls from the cosmological black holes.
Namely, the topological defect ceases to penetrate the cosmological black hole,
which is equivalent to the case that the domain wall's scalar field $X = 0$ over
the black hole event horizon. The previous considerations for {\it thin walls}
approximation indicate that for large mass black holes 
there is a solution with the wall intersecting black hole but
it gives no indication for small mass cosmological black holes.
\par
We shall give here the bound on cosmological black hole radius $r_{BH}$
for which the the domain wall scalar field $X$ must be expelled from the cosmological
black hole.
First,
let us suppose that there is a solution of the Eq. of motion for scalar
field that penetrates the horizon of black hole.
Thus, Eq.(\ref{xx}) on the horizon of the black hole becomes
\be
X_{,\theta \theta} + \cot \theta X_{,\theta} - r_{BH}^2
{X(X^2 - 1) \over w^2} = 0.
\label{hor}
\ee
As we take the third derivative of (\ref{hor}), then one obtains the following:
\be
X_{,\theta \theta \theta} = - \cot \theta~ X_{,\theta \theta}
+ X_{,\theta} \bigg[ {1 \over \sin^{2} \theta} + r_{BH}^{2} 
\bigg({3 X^2 \over w^2} - 1 \bigg) \bigg].
\label{thr}
\ee
The nontrivial solution satisfied the condition $X(\pi/2) = 0$ which
caused that $X_{,\theta\theta}(\pi /2) = 0$. Let us suppose that
$X_{,\theta}(\pi /2)>0$ so that for $X_{,\theta}$ one has maximum (minimum)
at ${\pi \over 2}$ for $r_{BH}^2>w^2$~$(r_{BH}<w^2)$, because from
Eq.(\ref{thr}) one can see that at any turning point of $X_{\theta}$
we get
\be
X_{,\theta \theta \theta} = X_{, \theta} \bigg[
{1 \over \sin^{2} \theta}+ r_{BH}^{2} \bigg({3 X^2 \over w^2} - 1 \bigg) \bigg]
- \cot \theta~ X_{,\theta \theta}>X_{,\theta} \bigg[
1 - r_{BH}^2 \bigg].
\ee
For $ r_{BH}^2<w^2$ any turning point of $X_{\theta}$ is a minimum
which is consistent with $X_{\theta} = 0$ at $\theta = 0, \pi$. Therefore
for $ r_{BH}^2<w^2$ the only possible solution is $X = 0$ on the horizon.
In this case one has the expulsion.
\par
Now we treat the second case when $ r_{BH}^2>w^2$. 
When a stable expelling solution takes place one has
that in the vicinity of the black hole horizon it monotonically relaxes to a kink
solution as we move away from the horizon. thus $X_{,\theta}$ is of the same sign
(without loss of generality we take that $X_{,\theta}>0$ for our condition).
Let us consider 
the situation
that in the vicinity of the black hole horizon one has $r = r_{BH} + \beta$,
where $\beta$ is a small parameter.
By virtue of this Eq. of motion takes the form as follows:
\be
X_{,\theta \theta} = - \cot \theta~ X_{\theta} - \bigg(
r_{BH}^2 g^{rr} X_{r} \bigg)_{,r} - r_{BH}^2 {X\over w^2} + {\cal O}(\beta).
\label{kap}
\ee
Consider next the higher derivative of scalar field $X$.
One can show that $X_{, \theta \theta},~ X_{,\theta \theta \theta} \ge 0$
and $X_{,\theta \theta \theta \theta} \leq 0$ on $[{\pi \over 2}, \pi]$
for $r_{BH}^2>w^2$. For since we have the following:
\ben
{\pi \over 4} 
X_{, \theta}\bigg( {\pi \over 2} \bigg)<X \bigg( \pi \bigg)<{\pi \over 2} X_{, \theta}
\bigg( {\pi \over 2} \bigg), \\
{\pi \over 4} 
\mid X_{, \theta \theta}\bigg( {\pi \over 2} \bigg) \mid<X_{, \theta} 
\bigg( \pi \bigg)<{\pi \over 2} \mid X_{, \theta \theta} \bigg( \pi \bigg) \mid.
\een
Having in mind relation (\ref{kap}) and the above inequalities we obtain
\be
{\pi^2 \over 16} {r_{BH}^2 \over w^2} X \bigg( \pi \bigg)<X 
\bigg( \pi \bigg)<{\pi^2 \over 4} {r_{BH}^2 \over w^2} X \bigg( \pi \bigg),
\ee
i.e., $ {2 w \over \pi}<w<r_{BH}<{4 w \over \pi}.$\\
Therefore for $r_{BH}>4 w / \pi$ there is no expulsion of scalar field
from the cosmological black hole interior.
\par
The next problem we shall consider is a
cosmological black hole sitting
inside the domain wall. Inside the domain wall the potential terms
are negligible comparing to the gradient ones. This fact justifies neglecting it.
We provide an ansatz for the scalar field in the form
$X(r, \theta) = b(r) \cos \theta$.
Plugging this ansatz into the scalar equation of motion gives
\be
{d \over dr} \bigg[ r^2 V(r) {d b \over dr} \bigg] - 2 b(r) = 0,
\label{thi}
\ee
where $V(r) = 1 - {2 M \over r} + {Q^{2} \over r^2} -{ \Lambda \over 3} r^2$.
To overcome the difficulty in solving Eq.(\ref{thi}) in terms of known functions
we shall approximate $V(r)$ near the cosmological black hole horizon by the relation
\be
V(r) \sim 2 \kappa_{BH} \big( r - r_{BH} \big),
\ee
where $\kappa_{BH}$ is the surface gravity of the black hole. Further, 
we define new variables
\be
\eta^2 = { r - r_{BH} \over 2 \kappa_{BH} r_{BH}^2}.
\ee
It can be easily verified that the relation (\ref{thi}) yields
\be
{d^2 b \over d \eta^2} + {1 \over \eta} {d b \over d \eta^2} - 8 b = 0.
\ee
The solution of it may be expressed in the form
\be
b = A I_{0}\bigg( 2 \sqrt{2} \eta \bigg) + B K_{0}\bigg( 2 \sqrt{2} \eta \bigg),
\ee
where $I_{0}, K_{0}$ are modified Bessel functions
and $A,B$ are constant.\\
Let us suppose that we have expulsion on the horizon
of the black hole and find the form of scalar
field in this case.
On the horizon of the black hole
one has $X \mid_{ r = r_{BH}} = 0$. Moreover, 
we know the asymptotic behaviour of the scalar field at infinity.
On this simplifications the scalar field can be brought to the form
as follows:
\be
X(r, \theta) \sim \sqrt{2} (r - r_{BH})^{1/4} e^{- \psi (r - r_{BH})^{1/2}} I_{0}
\bigg( \psi (r - r_{BH})^{1/2} \bigg) \cos \theta,
\ee
where $\psi = 2 /r_{BH} (\kappa_{BH})^{1/2}$.

\subsection{Nariai and anti-Nariai black hole domain wall systems}
Now we turn our attention to the electromagnetic extension of the Nariai solution
introduced by Bertotti and Robinson \cite{ber}. In Refs.\cite{man95,haw95}
it has been concluded that a limiting approach would take near extreme RN-dS solution into
the charged Nariai one. Here by the extreme solution one means the situation when the 
cosmological and the outer black hole horizons coincide.
The charge Nariai metric is given by \cite{dia03b}
\be
ds^2 = {{R_{0}}^2 \over K_{0}} \bigg(
- \sin^2 \ga d \tau^2 + d \ga^2 \bigg)
 + R_{0}^2 \bigg( d \theta^2  + \sin^2 \theta d \phi^2
\bigg),
\ee
where $\ga$ and $\theta$ both run from $0$ to $\pi$ and $\phi$ has the period $2 \pi$
while $R_{0}$ and $K_{0}$
are positive constants. $K_{0}$ satisfies the additional condition
$0<K_{0} \le 1$. The magnetic and electric charges are given respectively as
$F_{\theta \phi} = q$ and $F_{\tau \ga} = q/K_{0}$.
The following redefinition of the coordinates
\be
\sin^2 \ga = 1 - {K_{0} \over R_{0}} R^2, \qquad \tau = 
\sqrt{{K_{0} \over R_{0}^2}} T,
\ee
enables us to write the metric as
\be
ds^2 = - N(R) dT^2 + {dR^2 \over N(R)} R_{0}^2
\bigg( d\theta^2 + \sin^2 \theta d \phi^2 \bigg),
\ee
where $N(R) = 1 - {K_{0} \over R_{0}^2} R^2$. The cosmological constant $\La$
and the charge are related by the conditions
\be
\La = {1 + K_{0} \over 2 R_{0}^2}, \qquad
q^2 = {1 - K_{0} \over 2}R_{0}^2.
\ee
The equation of motion for
 $X$ field implies
\be
\p_{R} \big( N(R) \p_{R} X \big) + {1 \over R_{0}^2}X_{,\theta \theta} + 
{1 \over R_{0}^2}\cot \theta X_{, \theta} = {1 \over w^2}
X \big( X^2 - 1 \big).
\label{ext1}
\ee
Suppose, that we have a flux expulsion on the horizon and try to find
possible conditions for it occurs \cite{rog01}. Just, on the horizon we have
$X = 0$ and $N(R) \rightarrow 0$ near the black hole horizon while $R_{0}^2 X^3 \ll 1$. 
Having all these in mind and we integrate relation 
(\ref{ext1}) on the interval
$(\theta, {\pi \over 2})$, for $\theta>\beta_{0}$. 
It reveals the following:
\be
\p_{\theta} X(\theta)>{1 \over w^2}\cot \theta~ X~ R_{0}^2.
\ee
Next using the fact that $X_{,\theta \theta}<0$ on $[0, {\pi \over 2}]$
we deduce that 
\be
X_{,\theta \theta}<{X(\theta) - X(\theta_{0})  \over \theta - \theta_{0}}<{X(\theta)
\over \theta - \theta_{0}}<{X(\theta) \over \theta - \beta},
\ee
which in turn enables us to write the relation
\be
{1 \over R_{0}^2}>{1 \over w^2}\big( \theta - \beta \big) \cot \theta.
\label{in}
\ee
The above inequalities
must hold over the range of $\theta \in (\beta, {\pi \over 2})$ if
the expulsion takes place. Because of the fact that
$\theta - \beta>0$, $\cot \theta$ on this
interval is greater than zero, then the relation (\ref{in}) always holds and
one gets the expulsion of the thin domain wall from the charged Nariai 
black hole.
\par
Applying the same considerations to the anti-Nariai black hole which
the line element has the form \cite{dia03b}
\be
ds^2 = - N(R) dT^2 + {dR^2 \over N(R)} R_{0}^2
\bigg( d\theta^2 + \sinh^2 \theta d \phi^2 \bigg),
\ee
where $N(R) = - 1 + {K_{0} \over {R_{0}}^2} R^2$ and
\be
\La = - {1 + K_{0} \over 2 R_{0}^2}<0, \qquad
q^2 = {K_{0} - 1 \over 2}R_{0}^2,
\ee
the previous arguments can be repeated leading finally to the inequality
\be
{1 \over R_{0}^2}>{1 \over w^2}\big( \theta - \beta \big) \coth \theta.
\label{ina}
\ee
For the range of $\theta \in (\beta, {\pi \over 2})$ the relation always holds
revealing the fact  that as in charged Nariai case we have also the expulsion of 
the thin domain walls from charged anti-Nariai black holes.
\par
Summing it all we have derived a bound on the cosmological black
hole radius for which the expulsion occurred. Studying a cosmological
black hole sitting inside a domain wall we found the approximate solution
for the equation of motion for scalar field $X$.
We also showed that
both in the case of Nariai and anti-Nariai black holes
one had to do with
the complete expulsion of the domain wall's scalar fields from these black holes
(the so-called {\it Meissner effect}).

\section{Nucleation of cosmological black holes on domain walls}
In this section the problem of nucleation of cosmological black holes on the domain walls will
attract our attention. It turned out that most topological defects as strings, domain walls are
equipped with tension that caused them to be unstable to snapping or forming black holes on them.
One can say about two phenomena which can occur in the presence of domain wall, 
i.e., nucleating of a black hole on domain wall
or nucleation of black holes enclosed by the domain wall.\\
First we shall shed some light in the presence of domain wall on the process of forming the
cosmological black holes on domain walls. The probability of the process is given by
$exp[- (I - I_{0})]$, where $I_{0}$ is the Euclidean action of the initial
configuration. On the other hand, $I$ stands for the Euclidean action of the final state
with a domain wall and a cosmological black hole on it. The exponent in question
can be viewed as the ratio of probabilities of nucleation of cosmological black holes
on the domain wall. In order to build the domain wall-black hole instanton we shall follow the footsteps
described in \cite{haw96}. Then it follows that we have
\be
I = - {1 \over 4}\bigg( {\cal A}_{acc} + {\cal A}_{BH} \bigg),
\label{hh}
\ee
where ${\cal A}_{acc}$ is the area of cosmological (accelerated horizon), 
${\cal A}_{BH}$ is the area
of the black hole horizon. Consequently, Eq.(\ref{hh}) yields the result that
\be
I = - {2 \pi y_{4} \over A^2 \mid G'(y_{4}) \mid}
\bigg[
{1 \over (y_{3} - y_{4}) y_{3}} + {1 \over (y_{2} - y_{4}) y_{2}} \bigg].
\ee
The Euclidean action for a domain wall can be expressed as
$I_{0} = - 1/8 \pi \sigma^2$ \cite{cal96}.
The probability of this process will be more transparent 
in the limit of small black holes for which
$M, Q \ll 1/A$. The resulting result implies
\be
I = - {1 \over 8 \pi \sigma^2} + {M \over \sigma} 
+ {3 M k \over l^2 \sigma^2} + {\cal O}(A^2).
\ee
Thus, one gets
\be
I - I_{0} = (I - I_{0})_{RN-dw} + {3 M k \over l^2 \sigma^2},
\ee
where $(I - I_{0})_{RN-dw}$ describes the value of 
the exponent for nucleation rate of RN black hole domain wall
system.
Here the mass of a black hole is a parameter which can be varied independently of the
wall tension. This fact envisages that it can be arbitrary small.
\par
The problem of production of neutral and charged pairs of black holes was intensively
studied in literature \cite{cal96} while cosmological pairs black holes 
production was elaborated in Ref.\cite{man98}.
For the completeness of discussion and reader's convenience
we briefly comment on these results.
\par
The general situation comprises constructing a two-sided bubble by taking two 
regions of appropriate spacetime and joining them along the boundary satisfying the
Israel matching conditions. We shall pay attention to
nucleation of static black holes, i.e., black holes
which attractive gravitational energy exactly counterbalanced the repulsive
energy of domain walls. The matching conditions may be given by
$\sqrt{V(r) - \dot r} = 2 \pi \sigma r$. The above Eq. can be interpreted as 
the motion of a fictitious particle in a potential $v = V(r) - (2 \pi \sigma r)^2$.
On the other hand, the static 
solution for which $r = r_{st}$ has zero energy 
and 
it is obtained by solving the above matching 
condition under the supplementary condition of the form ${\p v \over \p r} = 0$.\\
The Euclidean action for the instanton is given by
\be
I_{E} = - {1 \over 16 \pi} \int_{M_{E}} d^4 x \sqrt{g} \bigg(
R - 2 \Lambda - F_{\mu \nu}F^{\mu \nu}\bigg)
- {\sigma \over 2} \int_{W} d^3 x \sqrt{h},
\label{pro}
\ee
where Euclidean section $M_{E}$ includes the values on both sides of the domain wall.
After some further reductions in which we make use of Einstein's Eqs.
relation (\ref{pro}) can be brought to the form
\be
I_{E} = \beta_{BH} \bigg[
- 2 \pi \sigma \sqrt{V(r_{st})} r_{st}^2 + Q^2 \bigg(
{1 \over r_{BH}} - {1 \over r_{st}} \bigg) -
{k \over \pi l^2} \bigg(
r_{st}^3 - r_{BH}^3 \bigg) \bigg],
\ee
where $\beta_{BH}$ is the black hole period in imaginary time $\tau = - it$.
The probability for the pairs creation of a static cosmological black holes
in the vicinity of a domain wall will be given by $exp(- I_{E})$.
\par
In contrast to the process of production of cosmological black holes on the domain walls
the nucleation in the presence of a domain wall is chacterized by the fixed 
mass \cite{man98,cha98} depending on the value of the domain wall tension $\sigma$.
For it cannot be varied independently of the domain wall's tension a certain
(large size) black holes can be nucleated. Hence, this process will be heavily suppressed.

\section{Conclusions}
In this paper we have studied the problem of cosmological black holes on the domain walls.
By means of the C-metric construction presented in \cite{emp00a,emp00b} we 
derived the metric for
the infinitisemally thin domain wall intersecting a cosmological black hole.
We also examined the problem of self-interacting cosmological domain wall and found the
approximated solution for the scalar field valid for large distances outside it.
We analyzed the scalar field equation of motion in the background of RN-AdS, RN-dS
and C-metric finding the {\it thin wall } approximation which was used in examining the
back reaction problem. As in general relativity \cite{emp01} and in dilaton 
gravity \cite{rog01,mod03} we justified that the 
the {\it singular} behaviour of the zero thickness domain wall
could be smoothed out by a core of a topological (thick) domain wall.
We analyzed the domain wall cosmological black hole system
and gave analytical
arguments for the domain wall's scalar fields to be expelled from the cosmological black hole.
Only for the charged Nariai and anti-Nariai black holes we analytically revealed
the expulsion of scalar fields (the analog of the {\it Meissner effect}).
Of course, there are preliminary analytic studies. Due to the complications of equations
of motion for scalar field
in other interesting backgrounds numerical studies should be conducted. 
Work in this direction is in progress
\cite{mod04}. 
\par
We compared two processes of nucleation of cosmological black holes on the domain wall
and the pairs productions of the black holes  in the presence of the domain wall.
It happened that the mass parameter of black holes on the domain walls could be carried
independently to the domain wall surface density thus could be arbitrary small. 
In contrast to this phenomenon the nucleation of pairs of cosmological black 
holes in the presence of domain wall is heavily suppressed due to the fact that
a certain large size black holes can be produced. The black hole's
mass depends on the domain wall tension. Summing it all up one 
can draw a conclusion that domain wall will prefer to nucleate
 small black holes on it rather than large one inside it.
The extension of these problems to the situations in which charged domain walls 
or supersymmetric domain walls and dilaton couplings are included remains an 
interesting open question. We hope to return to it elsewhere.

\vspace{3cm}
\noindent
{\bf Acknowledgements:}\\
M.R. was supported in part by KBN grant No. 2 P03B 124 24.

\end{document}